\documentclass[
twocolumn,
prl,unsortedaddress,showpacs,floatfix,citeautoscript,nofootinbib,reprint]{revtex4}
\usepackage{amsmath}
\usepackage{amssymb}
\usepackage{amsfonts}
\usepackage[T2A]{fontenc}
\usepackage{verbatim}
\usepackage{dcolumn}
\usepackage{tabularx}
\normalfont\sffamily\rmfamily\ttfamily

{}
\begin{document}

\title{Dynamics of solitary waves in ultracold gases
in terms of observable quantities}
\author{L. P. Pitaevskii$^{1,2}$} 
\affiliation{$^{1}$Kapitza Institute for Physical Problems RAS, Kosygina 2, 119334
Moscow, Russia\\
$^{2}$INO-CNR BEC Center and Dipartimento di Fisica, Universit\`a di Trento,
38123 Povo, Italy}

\begin{abstract}
A variety of solitary waves, such as solitons, vortex
rings, solitonic vortices, and more complex entities, have recently been predicted to exist. They can move in superfluid
ultracold gases along elongated traps. The theoretical description of this motion requires knowledge of the inertial soliton
mass and the effective number of particles in it as functions of
the soliton energy. While these functions can be calculated by a
microscopic theory, it is also possible to express them directly in
terms of observable quantities, such as the order parameter
phase jump and the particle number depletion in the soliton.
In this article, the corresponding equations are derived in a
simple and physically clear way and applied to the recently
predicted `magnetic soliton' in mixtures of Bose gases in various spin states. 
\end{abstract}
\pacs{03.75.Lm, 3.75.Kk, 67.85.De\\DOI: https://doi.org/10.3367/UFNe.2016.08.037891 }
\maketitle

\section{1. Introduction.
Macroscopic equations of motion
of a soliton and observable quantities}

The existence of solitary waves - solitons, vortex rings,
solitonic vortices, and more complex entities is an
important property of quantum gases. The detailed study of
the solitary waves has become possible with the development
of experimental techniques for confining atomic gases in
magnetic and optical traps and cooling them to ultralow
temperatures. At such temperatures, a gas becomes super-
fluid, and solitary waves can be observed as moving non-
decaying objects. (For brevity, in what follows I refer to
these waves as solitons.) The essential point is that these
excitations, while macroscopic, contain a certain highly
perturbed core region, which has a structure dependent on
the specific properties of the system and which should be
described in microscopic terms. A typical experiment for
studying such objects consists in observing their motion in
an elongated trap. If the trap is sufficiently long, the problem
can be solved in two stages, by first finding the solution for a
uniform cylindrical trap and then determining the motion of a
soliton in a finite trap applying macroscopic equations of
motion. It turns out, however, that soliton dynamics in such a
one-dimensional geometry exhibits interesting features due to
the occurrence of a jump in the order parameter phase (see
Section 2). For example, this jump substantially affects the
canonical momentum of the soliton, a fact which was already
noted in the classical work of T. Tsuzuki who used the Gross-Pitaevskii 
(below abbreviated GP) equations to construct a
plane \footnote{Plane in the sense that all the quantities involved are assumed to depend
on the $z$ coordinate alone.} soliton in a dilute Bose-Einstein condensed gas \cite{Ts71}.

The soliton equation of motion in an elongated trap can
be derived from the energy conservation law. A convenient
way to do this involves using the soliton velocity $V$ and the
chemical potential $\mu$ of the gas as independent variables. The
soliton energy should then be determined by employing the
thermodynamic potential of the grand canonical ensemble,
$E^{\prime }=E_{0}^{\prime
}-\mu N$, where $E$ is the energy, and $N$ is the number of
 particles in the system. For a soliton in a uniform trap, one
has $E^{\prime }=E_{0}^{\prime
}+$ $\epsilon \left( V,\mu \right) $ (where $E_{0}^{\prime }$ is the ground state potential)
and $\epsilon \left(
V,\mu \right) $ is, by definition, the soliton energy. For a weakly
nonuniform (i.e. very elongated in the $z$-detection) trap, the
energy can be approximated as 
\begin{equation}
\epsilon \left( V,\mu ,z\right) =\epsilon \left( V,\mu -V_{\mathrm{ext}}\left( z\right) \right)\; ,  \label{epsilonz}
\end{equation}
where $z$ denotes the soliton center coordinate, and $V_{\mathrm{ext}}\left( z\right) $ is
the trap potential. (This is what is known as the local density
approximation.) For a harmonic trap, the potential takes the
form $V_{\mathrm{ext}}\left( z\right) =M\omega _{z}^{2}z^{2}/2$. The soliton equation of motion is
determined by the energy conservation $\epsilon \left( V,\mu ,z\right) =const.$\cite{BA00,KP04}.
 Differentiating Eqn (\ref{epsilonz}) with respect to time and noting
that $dz/dt = V$, the equation of motion assumes the form
\begin{equation}
m_{I }\frac{\partial V}{\partial t}\mathbf{=-}N_{S}\frac{\partial V}{\partial
z}\; ,  \label{Eq}
\end{equation}
where 
\begin{equation}
m_{I }=\frac{1}{V}\left( \frac{\partial \epsilon }{\partial V}\right) _{\mu
},\ N_{S}=-\left( \frac{\partial \epsilon }{\partial \mu }\right) _{V}\; .
\label{mINS}
\end{equation}
Here, $m_{I }$ has the meaning of the soliton inertial mass, and $N_{S}$
is the effective number of atoms in the soliton \footnote{The quantity $m_{I}$ is also often called 'effevtive mass' and denoted $m^{\ast}$, $m_{s}$ or $m_{eff}$.}.   Thus, knowing
the energy (\ref{epsilonz}) allows the determination of the soliton motion
in the trap. However, the calculation of $\epsilon$ is a challenging
theoretical problem in itself, so it is of great interest whether
$m_{I}$ and $N_{S}$ can be expressed directly in terms of observable
quantities. It is this question with which this paper is
concerned.

I will assume that a superfluid gas is described by a
complex order parameter $\Psi$, whose phase $\phi$ defines the
superfluid velocity through the equation 
\begin{equation}
\mathbf{v}=\frac{\hbar }{M}\mathbf{\nabla }\phi \; .  \label{vphi}
\end{equation}
For a gas of bosons,$\Psi$ is the wave function of the condensate,
and $M =m$. For a gas of superfluid fermions, $\Psi$ is the
Ginzburg-Landau wave function, and $M=  2m$ because this
function is the wave function of superconducting pairs. Here,
$m$ is the mass of the atom. Far from the soliton, where the
hydrodynamics is valid, the atomic density flux is $\bf{j} = n \bf{v}$,
where $n$ is the gas number density. There are two observable
quantities in terms of which I will express $m_{I}$ and $N_{S}$. One is
the order parameter phase jump 
\begin{equation}
\Delta \phi =\phi \left( z=\infty \right) -\phi \left( z=-\infty \right) \; .
\label{deltaphi}
\end{equation}
It can be shown \cite{KP03} that $\phi (\pm \infty )$ is independent of $x$ and $y$. The
phase jump $\Delta \phi$ can be (but has not yet been) measured in an
interference experiment and arises in a natural way in the
numerical calculations of the soliton structure. The other
observable quantity is the  'number of depleted
atoms in the soliton', namely
\begin{equation}
N_{D}=\int_{-\infty }^{\infty }\left[ n_{1}(z)-n_{1\infty }\right] dz \; ,
\label{ND}
\end{equation}
where $n_{1}(z)$ is the one-dimensional (i.e., $x-$ and $y-$integrated)
atomic number density in the soliton core, and $n_{1\infty }=n_{1}\left( z=\pm \infty \right) $ is the unperturbed density. Usually, the quantity $N_{D}$ is negative, and the soliton is 'dark'. Hence, the term
'depleted'. The quantity $N_{D}$ can be calculated from the
observed density distribution in the soliton. Notice that
$N_{D}\rightarrow 0$ at $V\rightarrow c$, where $c$ is the speed of sound. Indeed, it
is only density perturbations of infinitesimal amplitude that
can travel at the speed of sound.

\section{2. Canonical momentum and inertial mass}

The question of the effective mass can be conveniently
addressed by first discussing the soliton's momentum. By
definition, the momentum is directed along the z-axis and is
equal to
\begin{equation}
P =m \int j_{z}d^{3}x\;.  \label{pdef}
\end{equation}
It can be verified that the integral in this equation converges
for large z and is therefore uniquely defined. It is a simple
matter to express $P$ in terms of $N_{D}$. Consider the soliton to be
in the reference frame in which it is at rest, and let the flux of
atoms in this frame be $\mathbf{j}^{\left( 0\right) }$. Because motion in this frame is
steady-state, $\mathbf{j}^{\left( 0\right) }$
is independent of time and the continuity
equation takes the form $\nabla \cdot \mathbf{j}^{\left(
0\right) }=0$. Integrating this equation
with respect to $x$ and $y$ gives $\partial _{z}\int
j_{z}^{\left( 0\right) }dxdy=0$. Thus,
  $\int
j_{z}^{\left( 0\right) }dxdy$ is independent of z and is equal to its value at
$z=\pm \infty $, namely $-n_{1\infty }V$, because in this frame the liquid
outside of the soliton moves with velocity $-V$. On the other
hand, the flux in the laboratory frame is, via the Galilean
transformation, $j_{z}=j_{z}^{\left( 0\right)
}+Vn$. Substituting these formulas
into Eqn (\ref{pdef}), we find (see Ref. \cite{SDPS11})  
\begin{equation}
P = m\int_{-\infty }^{\infty }\left[ n_{1}(z)-n_{1\infty }\right]
dzV=mN_{D}V \; .  \label{loc}
\end{equation}
I will call the quantity $P$ in (\ref{pdef})-(\ref{loc}) the `local
momentum'. The local momentum becomes clearly zero at
$V=  0$ and also goes to zero together with $N_{D}$ as the velocity $V$
approaches the speed of sound. The important feature of the
one-dimensional geometry considered is that $P$ is not identical
to the canonical momentum $P_{C}$ which satisfies the equation
\begin{equation}
\left( \frac{\partial \epsilon }{\partial P_{C}}\right) _{\mu }=V\;,
\label{PC}
\end{equation} 
or equivalently
\begin{equation}
m_{I}=\frac{1}{V}\left( \frac{\partial \epsilon }{\partial V}\right) _{\mu
}=\left( \frac{\partial P_{C}}{\partial V}\right) _{\mu }\;.  \label{mIPC}
\end{equation}
The physical meaning of the difference between $P$ and $P_{C}$ is
currently well understood (see, for example, monograph \cite{PS16},
Ch. 5). Let the soliton move in a trap folded into a toroidal
ring of radius $R$. Further, let the radius be so large that the
curvature of the trap has no effect on the soliton dynamics.
This implies, in fact, the `thermodynamic limit' for $R\rightarrow \infty $.
Upon creating a soliton in a toroidal trap, the wave function
should remain unique. The presence of a phase jump Df is in
itself a factor that violates this uniqueness. This means that
the creation of a soliton occurs with the appearance outside of
it of a counterflow which compensates for the phase jump.
The momentum of the counterflow makes a contribution to
the canonical momentum. The counterflow velocity is
calculated from the condition that
\begin{equation}
\Delta \phi _{cfl}=2\pi Rv_{cfl}M/\hbar =-\Delta \phi \;,  \label{Dfcfl}
\end{equation}
where account was taken of the fact that outside of the
soliton, i.e., for
\begin{equation}
\vert z \vert \gg \xi \;,  \label{Rxi}
\end{equation}
where $\xi$ is the soliton core thickness, the counterflow velocity
can be considered constant across the cross section of the
trap. Accordingly, the counterflow momentum is $P_{cfl}=2\pi Rmn_{1\infty
}v_{cfl}=-\hbar mn_{1\infty }\Delta \phi /M$. The canonical momentum
represents now the total momentum, i.e., equals the sum of
$P+P_{cfl}$\footnote{The question of how actually to calculate the momentum is not addressed 
in this article. For solitary waves comprising quantized vortices, an
equivalent equation derived in Ref. \cite{Pit14} is more suitable.}: 
\begin{equation}
P_{C}=mN_{D}V-\hbar n_{1\infty }\frac{m}{M}\Delta \phi \;. \label{PCDphi}
\end{equation}
Differentiating this equation with respect to $V$ and taking into
account Eqn (\ref{mIPC}), we arrive at the sought after relation among
$m_{I}$, $N_{D}$, and $\Delta \phi$:
\begin{equation}
m_{I}=m\left( \frac{\partial }{\partial V}N_{D}V\right) _{\mu }-\hbar
n_{1\infty }\frac{m}{M}\left( \frac{\partial }{\partial V}\Delta \phi
\right) _{\mu }\;.  \label{mINDDphi}
\end{equation}
This equation was used in Ref. \cite{SDPS11} when studying soliton
dynamics in a superfluid Fermi gas under unitarity conditions. 
The trapped soliton was assumed there to perform only
small oscillations, and the difference between $N_{S}$ and $N_{D}$ was
not significant (see Section 3).

\section{3. Effective number of particles
and depletion of particle number }

My task now reduces to expressing the effective number of
particles $N_S$ in a soliton in terms of $N_{D}$ and $\Delta \phi $. For a soliton
at rest, the problem is simple: the system is steady-state, so
that it is possible to apply the thermodynamic relation for the
energy $E^{\prime}$ of the grand canonical ensemble: $N=-\left( \partial E^{\prime
}/\partial \mu \right) $,
where $N$ is the number of atoms in the system. By definition,
the soliton energy is $E^{\prime }(\mu )=E_{0}(\mu
)+\epsilon \left( V=0,\mu \right) $, where
$E_{0}(\mu )$ is the energy in the absence of the soliton, and
$N=N_{0}+N_{D}$. Thus, we obtain
\begin{equation}
N_{D}=-\frac{\partial \epsilon (V=0,\mu )}{\partial \mu }=N_{S}\; .
\label{ND0}
\end{equation}
This formula, however, does not apply to a moving soliton,
because in the laboratory frame the problem is not
stationary, and so the question requires a special analysis.
Such an analysis has indeed been made in Refs \cite{SGK12,Cam13} which
used $\mu $ and $P_{C}$ as calculation variables. (The distinction
between $N_{D}$ and $N_{S}$ was brought to my attention by
D~M~Gangardt.) My hope is that the derivation below will
give insight into the physical meaning of the equations
obtained. The investigation of this interesting question has long
been hindered by the fact, to some extent accidental, that
for a plane Tsuzuki soliton the equality $N_{D} = N_{S}$ is also valid
at a finite velocity.

In order to apply thermodynamics to a moving soliton, it
is necessary to transfer to a reference frame which moves
relative the lab frame with velocity V and where the soliton is
at rest. According to the general Galilean transformation rule
for mechanical quantities, the soliton energy in this frame is
given by
\begin{equation}
\tilde{\epsilon}(V,\mu )=\epsilon (V,\mu )-P_{C}(V,\mu )V.  \label{etil}
\end{equation}—оответственно, изменение числа частиц при рождении солитона, аналогично (\ref{ND0}), равно\begin{equation}
\Delta N=-\frac{\partial \tilde{\epsilon}(V,\mu )}{\partial \mu }=-\frac{\partial }{\partial \mu }\left( \epsilon -P_{C}V\right) _{V}\ .  \label{DNV}
\end{equation}
This quantity cannot, however, be identified so far with $N_{D}$.
Due to the presence of the counterflow, also outside of the
soliton the number of particles changes by a finite amount
$\Delta N_{cfl}$. As $R\rightarrow \infty $, the density of these particles tends to zero
and cannot be measured, so DNcfl should be subtracted from
$\Delta N$:
\begin{equation}
N_{D}=\Delta N-\Delta N_{cfl}\;.  \label{NDfin}
\end{equation}
The quantity $\Delta N_{cfl}$ is easily calculated using the Bernoulli
hydrodynamic equation valid in the superfluid gas outside of
the soliton:
\begin{equation}
\mu _{l}(n_{1\infty })+m\frac{v^{2}}{2}=\mu _{0}\;,  \label{mv}
\end{equation}
where $\mu _{l}(n_{1\infty })$
is the chemical potential of the gas at rest
expressed in terms of its number density, and $v$ is the gas
velocity. In the ground state, one has $v = -V$. Upon creating
a soliton, $v=-V+v_{cfl}$. Thus, the creation of a soliton
changes $\mu _{l}$ by the quantity $\Delta \mu _{l}=m\left[
\left( -V+v_{cfl}\right) ^{2}-V^{2}\right] /2\approx mVv_{cfl}$. The corresponding change in the number density is
$\Delta n_{1\infty }=\left( d\mu
/dn_{1\infty }\right) mVv_{cfl}$, and a change in the number of
particles is written out as
\begin{equation}
\Delta N_{cfl}=\frac{dn_{1\infty }}{d\mu }mV2\pi Rv_{cfl}=-\frac{dn_{1\infty
}}{d\mu }V\frac{m}{M}\Delta \phi \;,  \label{DNcfl}
\end{equation}
where I have made use of equation (\ref{Dfcfl}) for $2\pi
Rv_{cfl}$. Substituting Eqns (\ref{DNV}) and (\ref{DNcfl}) into Eqn (\ref{NDfin}, we finally
obtain the deficiency in the number of particles in the soliton:
\begin{equation}
N_{D}=-\frac{\partial }{\partial \mu }\left( \epsilon -P_{C}V\right) _{V}+\frac{dn_{1\infty }}{d\mu }V\frac{m}{M}\Delta \phi \; .  \label{NDfin1}
\end{equation}
Using definition (\ref{mINS}) of the effective number of atoms and
expression (\ref{PCDphi}) for momentum $P_{C}$, the desired expression of
$N_{S}$ in terms of $N_{D}$ and $\Delta \phi $ looks as follows:
\begin{equation}
N_{S}=N_{D}-mV^{2}\left( \frac{\partial }{\partial \mu }N_{D}\right)
_{V}-V\hbar n_{1\infty }\frac{m}{M}\left( \frac{\partial }{\partial \mu }\Delta \phi \right) _{V}.  \label{NSNDDf}
\end{equation}
For the Tsuzuki soliton, all the considered quantities can be
expressed explicitly in analytical form by solving the Gross-Pitaevskii 
equation. Equations ((\ref{mINDDphi}), (\ref{NDfin1}) then give
\begin{equation}
N_{S}=N_{D}=m_{I}/2  \label{GP}
\end{equation}
independent of velocity $V$. The same result is obtained by
directly integrating (\ref{ND}).

Let us return to (\ref{NDfin1}). We first eliminate $\Delta \phi $ from this
equation by making use of Eqn (\ref{PCDphi}) and then proceed by
using the following relation between the derivatives, which is
valid if canonical equation (\ref{PC}) is satisfied:
\begin{equation}
\frac{\partial }{\partial \mu }\left( \epsilon -P_{C}V\right) _{V}=\left( 
\frac{\partial \epsilon }{\partial \mu }\right) _{P_{C}}.
\end{equation}
As a result, we arrive at the equation for $N_{D}$:
\begin{equation}
\left( 1-\frac{mV^{2}}{n_{1\infty }}\frac{dn_{1\infty }}{d\mu }\right)
N_{D}=-\left[ \left( \frac{\partial \epsilon }{\partial \mu }\right)
_{P_{C}}+\frac{V}{n_{1\infty }}\frac{dn_{1\infty }}{d\mu }P_{C}\right] ,
\label{Brand}
\end{equation}
or, taking into account the equation for the speed of sound,
$c^{2}=\left( n_{1\infty
}/m\right) \left( d\mu /dn_{1\infty }\right) $, we find in the upshot:  
\begin{equation}
N_{D}=-\frac{\left( \partial \epsilon /\partial \mu \right)
_{P_{C}}+VP_{C}/mc^{2}}{1-V^{2}/c^{2}}.  \label{NDc}
\end{equation}
It is in this form that the equation was first presented in
Ref. \cite{Cam13}, see Eqn (3.16). This equation is convenient to apply
when the soliton energy is expressed as a function of $P_{C}$, not
of $V$, as is commonly the case with microscopic theories based
on exactly solvable models. As for equations, analogous to the
mean field theory GP equation, calculations using them
involve the variables $\mu$, $V$. Equation (\ref{Brand}) has been employed
to calculate $N_{D}$ in a one-dimensional Fermi gas \cite{SB16}. Notice
that by substituting $N_{D}$ from Eqn (\ref{NDc}) into Eqn (\ref{PCDphi}) we
obtain, by resorting to the canonical momentum definition
(\ref{PC}), the equation which defines the phase jump $\Delta \phi $ in terms of
the derivatives of $\epsilon
(P_{C},\mu )$.

\section{4. Nontrivial example. Magnetic soliton}

As I have already noted, the application of the obtained
relations to a plane soliton in a weakly nonideal Bose gas gives
a trivial result (\ref{GP}). There are currently only few analytical
solutions that describe solitons. In this section, I will apply the
obtained relations to the recently predicted phenomenon of a
`magnetic soliton' in the mixture of two Bose-Einstein
condensates residing in different hyperfine triplet states \cite{QPS16}.
Such a mixture is described by a system of two coupled
Gross-Pitaevskii type equations. However, the problem
greatly simplifies if the inequality
\begin{equation}
\delta а \equiv a-a_{12}\ll a  \label{M1}
\end{equation}
holds, where $a=\sqrt{a_{11}a_{22}}$ and $a_{11}\approx a_{22}$
are scattering lengths
for atoms in the same state, and $a_{12}$ is the scattering length for
atoms in different states. Under this condition, the equations
describing the dynamics of the total gas number density $n=n_{1}+n_{2}$
 separate from those for the difference $n_{1}-n_{2}$ of
the component number densities. I will consider a symmetric
case, in which $a_{11}=a_{22}$ and the unperturbed densities are
equal, $n_{1}=n_{2}$. The essential point is that small perturbations
of the density $n$ travel at sound speed $c=\sqrt{gn/m}$,
 where $g$ is
the interaction constant equal to $g=4\pi \hbar ^{2}a/m$ in the case of
purely one-dimensional motion. The chemical potential is
equal to the usual value for a condensate, $\mu =gn$. As for the
difference $n_{1}-n_{2} $, which has the meaning of the gas spin
polarization, its perturbations travel with the speed of `spin
sound':
\begin{equation}
c_{s}=\sqrt{\alpha \frac{gn}{m}}=\sqrt{\alpha \frac{\mu }{m}},  \label{cs}
\end{equation}
where $\alpha =\delta a/2a\ll 1$ is the small parameter of the problem,
$c_{s} \ll c $. Notice that the total number density $n$ is
perturbed weakly. In a similar way, it is possible to construct
an analytical solution describing a plane `magnetic soliton',
i.e., a localized region in which $n_{1}-n_{2}\neq
0 $ , and the total number density $n$ is constant in the first approximation. The
variables involved are $n_{1}-n_{2}$ и фазы $\phi _{1}$
and $\phi _{2}$ of two
order parameters, instead of which it is convenient to
introduce the phase difference $\phi _{A}=\phi _{1}-\phi _{2}$ and the phase
sum $\phi_{B}=\phi _{1}+\phi _{2}$.  In the following, we will only need the
energy of the magnetic soliton \cite{QPS16}:
\begin{equation}
\epsilon _{M}=n\hbar \sqrt{c_{s}^{2}-V^{2}}  \label{em}
\end{equation}
and the jump in the phase sum $\phi _{B}:
$\begin{equation}
\Delta \phi _{B}=-2\arccos \left( \frac{V}{c_{s}}\right)\; .  \label{DfB}
\end{equation}
Equation (\ref{PCDphi}) for the canonical momentum is readily
extended to the case of two condensates:
\begin{equation}
P_{MC}=mN_{MD}V-\hbar \frac{n}{2}\Delta \phi _{B}.
\end{equation}
At this point, we should take into account that the total
number density $n$ also remains constant within the soliton,
which means that the first term in this equation is small and
can be disregarded. 
(It can be shown that $\left\vert N_{MD}\right\vert V\sim \alpha \hbar n\Delta \phi_{B}$.) Thus, we obtain
\begin{equation}
P_{MC}\approx -\hbar \frac{n}{2}\Delta \phi _{B}=-\hbar n\arccos \left( 
\frac{V}{c_{s}}\right)\; .  \label{PMC}
\end{equation}
The canonical momentum of the magnetic soliton is almost
completely determined by the counterflow. From Eqn (\ref{PMC}),
the following simple expression for the soliton energy as a
function of the momentum can be derived:
\begin{equation}
\epsilon _{M}=n\hbar c_{s}\left\vert \sin \left( \frac{P_{MC}}{\hbar n}\right) \right\vert ,-\frac{\pi \hbar n}{2}\leqslant P_{MC}\leqslant \frac{\pi \hbar n}{2}\; .
\end{equation}
Differentiating expressions (\ref{em}) yields the soliton inertial
mass $m_{MI}$ and the effective number of particles $N_{MS}$:
\begin{eqnarray}
m_{MI}&=&-\frac{n\hbar }{\sqrt{c_{s}^{2}-V^{2}}}\; , \nonumber \\
N_{MS}=&-&\frac{\hbar }{g}\left[
\sqrt{c_{s}^{2}-V^{2}}+\frac{c_{s}^{2}}{2\sqrt{c_{s}^{2}-V^{2}}}\right] \; .
\label{mIM}
\end{eqnarray}
While both these quantities diverge on approaching the
magnetic sound speed, their ratio (which enters the equation
of motion (\ref{Eq})) remains finite:
\begin{equation}
\frac{N_{MS}}{m_{MI}}=\frac{1}{\mu }\left[ \frac{3}{2}c_{s}^{2}-V^{2}\right]
\; .  \label{NsM}
\end{equation}
Let us now evaluate the depletion of the number of particles
$N_{MD}$ in a magnetic soliton. Equation (\ref{NDfin1}) for the case of two
condensates takes the form
\begin{equation}
 N_{MD}=-\frac{\partial }{\partial \mu }\left( \epsilon -P_{C}V\right) _{V}+\frac{dn_{1\infty }}{d\mu }V\frac{m}{2}\Delta \phi \; , 
\end{equation}
or, expressing $P_{C}$ in terms of $\Delta \phi $ according to Eqn (\ref{PMC}), we
obtain
\begin{equation}
N_{MD}=-\frac{\partial \epsilon }{\partial \mu }-nV\frac{m}{2}\frac{\partial 
}{d\mu }\Delta \phi \; .
\end{equation}
A simple calculation gives
\begin{equation}
N_{MD}=-\frac{3\hbar }{2g}\sqrt{c_{s}^{2}-V^{2}}\; .
\end{equation}
We see that, unlike the Tsuzuki soliton, $N_{MS}$ and $N_{MD}$ behave
totally differently as $V\rightarrow c_{s}$. At $V = 0$, these quantities are
equal (as they should be).

\subsection{5. Derivation the relation between \protect{$N_{S}$}
and \protect{$N_{D}$} directly from equations}

In this section, we will derive equation (\ref{NDfin1}) directly from an
equation that generalizes the Gross-Pitaevskii equation.
Such generalized equations are widely applied in soliton
theory, if the system can be described by a mean field
theory. Among such problems is, in particular, that of
applying the GP equation to the study of solitary waves in a
cylindrical trap. The point is that simple relation (\ref{GP}) is valid
only for a plane Tsuzuki soliton. If, however, a radial
confining potential is present, then even in the case of an
ordinary soliton the equations obtained above should be
used, let alone more complex solitary waves that exist in
such a geometry \cite{KP03,BR02}.

I will assume that the system is described by the energy
functional  $H[\Psi ,\Psi ^{\ast }]$, where $\Psi$ is
the order parameter (or the wave function)
$\Psi (t,\mathbf{r})=\sqrt{n}e^{i\phi }$, so that $\mathbf{j=}n\frac{\hbar }{m}\nabla
\phi $. (I set  $M=m$, thus limiting the problem to the case of
bosons.)

The equation for $\Psi $ is obtained by varying of the the energy
functional of the grand canonical distribution $H^{\prime }[\Psi ,\Psi ^{\ast }]$
over $\Psi ^{\ast }$ 
\begin{equation}
i\hbar \partial _{t}\Psi =\frac{\delta H^{\prime }}{\delta \Psi ^{\ast }}\; ,
\end{equation}
where $H^{\prime }[\Psi ,\Psi ^{\ast }]$ is
\begin{equation}
H^{\prime }[\Psi ,\Psi ^{\ast },\mu ]=H[\Psi ,\Psi ^{\ast }]-E_{0}^{^{\prime
}}-\mu \int \left\vert \Psi \right\vert ^{2}d^{3}x\; .  \label{H'}
\end{equation}
For a running soliton, the solution is $\Psi =\Psi _{S}(z-Vt)$, so that
the equation can be rewritten as
\begin{equation}
-i\hbar V\partial _{z}\Psi _{S}=\frac{\delta H}{\delta \Psi ^{\ast }}\; .
\label{GPV}
\end{equation}
Let us now determine $N_{S}$ and $N_{D}$ in terms $H^{\prime }$. The energy of
the gas in the presence of a soliton is $\varepsilon =H^{\prime }[\Psi
_{S},\Psi _{S}^{\ast },\mu ]$ and $N_{S}=-\frac{d}{d\mu }H^{\prime
}[\Psi _{S},\Psi _{S}^{\ast },\mu ]$ .
On the other hand, differentiating Eqn (\ref{H'}) at constant 
obviously gives
\begin{equation*}
\frac{d}{d\mu }\left( H^{\prime }\right) _{\Psi }=-\int \left( \left\vert
\Psi \right\vert ^{2}-n_{\infty }\right) d^{3}x\; ,  
\end{equation*}
because $dE_{0}^{\prime }/d\mu =-N$ .
Thus, we obtain 
\begin{equation*}N_{D}=-\frac{d}{d\mu }\left( H^{\prime }\right) _{\Psi =\Psi
_{S}}\; .
\end{equation*}
Differentiation of the functional $H^{\prime }[\Psi
_{S},\Psi _{S}^{\ast },\mu ]$ yields
\begin{equation}
N_{S}=N_{D}-\int \left( \frac{\delta H^{\prime }}{\delta \Psi ^{\ast }}\frac{d\Psi ^{\ast }}{d\mu }+c.c.\right) d^{3}x\; ,
\end{equation}
where the replacement $\Psi \rightarrow \Psi_{S} $ should be made after differentiation. Eliminating the variational derivatives
with the help (\ref{GPV}) 
results in 
\begin{equation}
N_{S}=N_{D}-2\hbar V\rm{Im}
\int \left( \partial _{z}\Psi _{S}\frac{d\Psi%
_{S}^{\ast }}{d\mu }\right) d^{3}x \; .  \label{NSNDGP}
\end{equation}

Notice that, given this equation, no additional calculations
are needed to show that for a plane Tsuzuki soliton $N_{S}=N_{D}$.
As shown in Ref. \cite{PS16}, Eqn (5.56)] the solution in this case can
be written out in a form in which $\rm{Im}\Psi _{S}$ is independent of $z$
and $\mu$, bringing the second term in Eqn (\ref{NSNDGP}) to zero.
Substituting $\Psi _{S}=\sqrt{n}e^{i\phi }$ into Eqn (\ref{NSNDGP}) gives, after
simple algebra, the following expression
\begin{equation}
N_{S}-N_{D}=-\hbar V\int \left[ \frac{d}{d\mu }\left( n\partial _{z}\phi
\right) -\partial _{z}\left( n\frac{d\phi }{d\mu }\right) \right] d^{3}x.
\end{equation}

In the second term of this equation, we can integrate with
respect to $z$ to obtain 
\begin{equation*}
\int
\partial _{z}\left( n\frac{d\phi }{d\mu }\right) d^{3}x=n_{1\infty }\frac{d\Delta \phi }{d\mu }\;,
\end{equation*}
where account was taken of the fact that  $\phi (z\rightarrow \pm
\infty )$ is independent of
 $x$ and $y$. Using the local momentum definition (\ref{pdef}), we arrive at the equation 
\begin{equation}
N_{S}-N_{D}=-V\frac{dP}{d\mu }+n_{1\infty }V\hbar \frac{d\Delta \phi }{d\mu }\; ,
\label{NDfin2}
\end{equation}
which taking into account Eqn (\ref{PCDphi}), is identical to Eqn (\ref{NDfin1}).

\section{6. Conclusion}
The theoretical description of solitary waves is complex and
requires a microscopic theory, because the characteristic core
thickness is on the order of the correlation length of particles
in the system. Accordingly, the comparison of theory with
experiment provides an important insight into the validity of
the theoretical approach used. The study of soliton motion in
an elongated trap is the most important of the experiments
available. Comparing theory with experiment using the
observables $N_{D}$ and $\Delta \phi$ opens new possibilities. This is
especially true of numerical simulations, when these two
quantities are easily determined and comparison with the
obtained equation of motion makes it possible to test the self-consistency 
of calculations and their underlying theory, as
well as allowing the estimation of the computation accuracy.\\
\bigskip

\textbf{AFTERWORD}\\
 The present paper has been submitted to the
special October 2016 issue of \textit{Uspekhi Fizicheskikh Nauk}
[\textit{Physics-Uspekhi}] journal on the occasion of the 100th
anniversary of the birth of V.~L.~Ginzburg. He was
a remarkable physicist recognized for his classical contributions 
to diverse areas of theoretical physics. My personal
communications with him contributed much to my development
 as a scientist, and I was always an admirer of his
personal qualities. The beginning of our friendship - one
that lasted uninterrupted until his death - was our joint work
on the theory of superfluidity in $^{4}$He near the $\lambda$-point \cite{GP58, G04}.
From the very beginning I was greatly impressed by the
clarity of his thought and by the total absence of pomposity in
his manners. A second intersection of our activities had to do
with the theory of van der Waals forces. The theory of these
forces, which I E Dzyaloshinskii, E M Lifshitz, and I [15, 16]
developed for an electromagnetic field in energy-absorbing
dielectrics, was formulated in terms of quantum field
theoretical diagrammatic techniques. Ginzburg asked me then
whether this could be done without applying this technique,
and I was firmly confident it could not. Shortly thereafter,
however, he devised a clever way to overcome absorption-
related difficulties and developed (with Yu S Barash) a
complete theory [17] which, among other things, made it
possible to get rid of some cumbersome calculations. The
theory of solitons, which is the subject of this article, was to
my knowledge never among his interests. But the theory of
solitons in superfluid Fermi systems always and inevitably
uses the boson order parameter, i.e., the Ginzburg - Landau
wave function. It should be noted, by the way, that the
Ginzburg - Landau theory only increases in importance as
time goes on. Of course and alas, nothing can replace direct
personal communication – but a soothing fact is that
V L Ginzburg's work remains and will be of service to future
generations of physicists.
Of special note is V L Ginzburg's role as the Chief Editor
of the Uspekhi Fizicheskikh Nauk [Physics - Uspekhi] journal
[18]. His continual work in the journal lasted, literally, to the
last day of his life, and it was V L Ginzburg's constant effort
to pursue impartiality and firm principles in the editorial
policy.\\
\bigskip

\textbf{Acknowledgments}\\
I am grateful to D M Gangardt, F Dalfovo, and S Giorgini
for fruitful discussions of the subject. This work was
supported by ERC through a QGBE grant, by a QUIC
grant from the Horizon2020 FET program, and by Provincia
Autonoma di Trento.

\end{document}